\documentclass[aasms4,usenatbib]{mn2e}
\usepackage[paperwidth=20.99cm,paperheight=29.70cm]{geometry}
\usepackage{epsfig}
\usepackage{graphicx}
\usepackage{amssymb}
\usepackage{numprint}
\newcommand{\D}{\mathrm{d}}
\newcommand{\mr}[1]{\mathrm{#1}}
\newcommand{\R}[1]{\ensuremath{R_\mathrm{#1}}}
\newcommand{\T}[1]{\ensuremath{t_\mathrm{#1}}}
\newcommand{\Period}[1]{\ensuremath{P_\mathrm{#1}}}
\newcommand{\w}[1]{\ensuremath{\tau_\mathrm{#1}}}
\newcommand{\kmps}{\ensuremath{\mathrm{km\ s^{-1}}}}

\begin{document} 

\title[Evolution of isolated neutron stars till accretion]
{Evolution of isolated neutron stars till accretion. The role of initial magnetic field}

\author[P.A. Boldin, S.B. Popov]
{P.A. Boldin $^{1}$, S.B. Popov $^{2}$ 
\thanks{E-mail: polar@sai.msu.ru (SBP)}\\
$^1${\sl National Research Nuclear University "MEPhI",
Kashirskoe shosse 31, Moscow, 115409, Russia}\\
$^2${\sl Sternberg Astronomical Institute, Universitetski pr. 13,   
Moscow, 119991, Russia} }
\date{Accepted ......  Received ......; in original form ......
      }

\maketitle

\begin{abstract}
 We study evolution of isolated neutron stars on long time scale and calculate distribution of
these sources in the main evolutionary stages: Ejector, Propeller, Accretor, and Georotator. We compare different initial
magnetic field distributions taking into account a possibility of magnetic field decay,
 and include in our calculations the stage of subsonic Propeller.  

It is shown that though the subsonic
propeller stage can be relatively long, initially highly magnetized neutron stars ($B_0\ga 10^{13}$~G) reach the
accretion regime within the Galactic lifetime if their kick velocities are not too large. 
The fact that in previous studies made $>$10 years ago, such objects were not considered results in
a slight increase of the Accretor fraction in comparison with earlier conclusions. 
Most of the neutron stars similar to the Magnificent seven are expected to become accreting
from the interstellar medium after few billion years of their evolution. They are the main predecestors
of accreting isolated neutron stars.
\end{abstract}

\begin{keywords}
stars: neutron --- pulsars: general 
\end{keywords}

\section{Introduction}

 Accreting isolated neutron stars (AINS) were predicted  40 years ago
by \cite{s71} 
and independently by \cite{ors70}. 
In early 90s there was some enthusiasm due to the launch
of the ROSAT satellite, which was expected to find many sources of this kind \citep{tc91}. 
Several populational studies have been made \citep{blaes90,br91,bm93,mb94,blaes95,manning96}.
However, it came out that AINS, if they exist, are very elusive (Colpi et al. 1998). The main reason is that initial (kick)
velocities of NSs appeared to be significantly larger, than it have been thought before \citep{ll94}. 
Initial guess that the number of Accretors is small due to low luminosity of high-velocity NSs was shown to be wrong.
In a detailed study by \cite{census2000} 
(hereafter Paper I) it was shown that INS with Crab-like initial parameters
and constant magnetic fields
spend all their lives as Ejectors (we follow the classification summarized in \citealt{l92}
) if their initial velocities
are $\ga 100$~km~s$^{-1}$.  Then, the fraction of Accretors was mainly determined by the fraction of
low-velocity NSs.

 Up to the very end of 90s, it was believed that the wast majority of NSs are born similar to the Crab pulsar.
I.e., that they have short initial spin periods (from milliseconds to few tens of millisecond) and magnetic fields 
$B\sim 10^{12}$~G. Now it is believed, that about one half of NSs have different initial properties \citep{ptp2006,keane2008}.
There are at least three groups of sources with distinct parameters: compact central objects (CCOs) in
supernova remnants (SNR), magnetars (anomalous X-ray pulsars - AXPs, and soft gamma-ray repeaters - SGRs), and
cooling radioquiet NSs dubbed the Magnificent seven (M7) \citep{zoo08}.
CCOs have low initial fields $\sim 10^{11}$~G \citep{h07,gh2009}
and relatively long spin periods (hundreds of millisecond).
AXPs and SGRs have large fields $\sim 10^{14}$~G (see a review in \citealt{Mereghetti2008}). 
The Magnificent seven-like NSs have fields slightly above $10^{13}$~G \citep{haberl2007, kaplan2008}. 
Probably, some of rotating radio transients (RRATs, \citealt{rrats2006}) are similar to the M7. 
This variety in initial properties deserves new studies of evolving NSs using the population synthesis technique 
(see a review in \citealt{pp2007}).
In this paper we present the first step.

We describe two models. At first, we discuss a simple semianalytical approach, which is used to illustrate the main 
features of the scenario.  In this model velocities and ambient densities are not changing.
Then we present a detailed numerical model, which takes into account spatial movements of  
NSs in the Galactic potential and realistic
3D distribution of the interstellar medium (ISM).
Our main results are based on this model.
 
In the next section we present basic concepts used in both models, and describe each of them.
Then, in Sec. 3, we present results. Discussion is given in Sec.4. In the last section we present our conclusions.

\section{Models}

In this section we describe our models. We start with explanation of some basic processes and parameters of magneto-rotational
evolution used in both models. Then we discuss the semianalytical and the full numerical model, consequently.

\subsection{Basic processes and parameters}

Here we describe some aspects of magneto-rotational evolution implemented in both semianalytical 
and numerical models.

\subsubsection{Standard magneto-rotational evolution}

Here we mainly follow the approach described in \cite{l92}.
We consider a NS being born as an Ejector.
At this stage a relativistic wind and Poynting flux are so strong that
they prevent incoming matter to penetrate inside neither gravitational capture
radius, \R{G}, nor inside the light cylinder radius, \R{l}.
\R{G} represent the typical scale  at which the ISM is captured by the NS gravity:
\begin{equation}
\R{G} = 2GM/v_\mr{rel}^2,
\end{equation}
where $M$ is a NS mass and $v_\mr{rel}$ is a relative velocity of
the NS and the ISM.
The light cylinder radius is defined as:
\begin{equation}
\R{l} = c/\omega,
\end{equation}
where $\omega$ is the spin frequency of a NS.
The ejected matter creates a cavern in the ISM within the distance of the Shvartsman radius, \R{sh}, 
at which the 
magneto-dipole pressure, $P \sim \mu^2 / \R{l}^4R_\mathrm{Sh}^2$, is equal to the
ram pressure of the ISM, $P \sim \rho v_\mathrm{rel}^2$.
At this stage a young NS can be visible as a radiopulsar (PSR), and we assume that it
losses energy via relativistic wind
and Poynting flux according to the magneto-dipole formula:
\begin{equation}
\label{magneto-dipole-losses}
\frac{1}{2} \frac{\D I \omega^2 }{\D t} = - \frac{2}{3} \frac{\mu^2\ \omega^4}{c^3}\sin^2 \chi. 
\end{equation}
Here 
$I = 10^{45}$ g~cm$^2$ is a moment of inertia of a NS,
$\mu$    is a magnetic dipole moment, 
$\chi$   is an angle between rotational and magnetic axis,
   which is assumed to be ${\pi}/{2}$ everywhere below.

The Ejector stage finish when the Shvartsman radius, $\R{sh}$, becomes less than $\R{G}$.
The regime changes because after matter appear inside  \R{G}, its pressure start to grow
 $P_\mathrm{matter} \sim r^{-5/2}$. This more rapidly than the relativistic wind pressure growth:
$P_\mathrm{wind}  \sim r^{-2}$. So, the condition $P_\mathrm{matter} >
P_\mathrm{wind}$ is reached and the pulsar wind cannot stop the incoming flow.

Another reason that causes the Ejector stage to cease is disappearence
of the magneto-dipole emission. This happens when \R{sh} 
becomes less than the light cylinder radius. So, now matter fills the
light cylinder, preventing the creation of the magneto-dipole emission.

For isolated NSs both cases can happen not only due to spin-down,
but also because an object enters a more dense region of the ISM. 
Variations of the relative velocity of a NS and the ISM 
are important, too.
For small velocities $\R{G}$ can become larger than  $\R{sh}$.
Oppositely, for large velocities $\R{sh}$ can become smaller 
than the light cylinder radius due to the ram pressure
(in this case $\R{G} < \R{l}$ due to a large velocity).

To summarize, when at least one of conditions $\R{sh} < \R{G}$ or $\R{sh} < \R{l}$
is met, the Ejector stage ceases.
After that, matter falls down till its pressure is counterbalanced by 
the magnetic field pressure. The radius at which these pressures are equalized
is called the Alfven radius:
\begin{equation}
\R{A}=\left( \frac{\mu^2}{2 \dot M \sqrt{2GM}} \right)^{2/7}
\end{equation}
for the case $\R{A} < \R{G}$ and
\begin{equation}
\R{A}=
    \left(
	\frac{2\mu^2 G^2 M^2}{\dot{M}v_\mr{rel}^5}
    \right)^{1/6}
\end{equation}
for the opposite. $\dot M$ is the accretion rate.

The Propeller stage begins if  $\R{A} > \R{c}$, where 
\begin{equation}
\R{c} = \left(\frac{GM}{\omega^2}\right)^{1/3}
\end{equation}
is the corotation radius at which the solid body rotational velocity equals the escape
velocity.
At this stage matter is ``propelled'' away from a star, 
because of its interaction with the magnetosphere. 

For the propeller stage we use the model proposed by \cite{shakura75}.
The same approach was also used in Paper I\nocite{census2000}.
The period derivative can be written as:
\begin{equation}
\label{P_spindown}
\frac{\D P}{\D t} = \dot{M} \R{A}^2 P 
I^{-1} \simeq K P^\alpha\ \mr{ s\,s^{-1} }.
\end{equation}
For $\alpha = 1$ we obtain:
\begin{equation}
K = 2.4\times10^{-14} \mu_{30}^{8/7} n^{3/7}v_{10}^{-9/7}$~s$^{-1},
\end{equation}
where
$\mu_{30}$ is a magnetic dipole moment in units of $10^{30}$ G cm$^{-3}$,
$n$ is the ISM number density,
$v_{10}$ is the total velocity $\sqrt{ a_\mr{s}^2 + v_\mr{rel}^2 }$
	in units of 10 km~s$^{-1}$, $a_\mr{s} = 10\ \kmps$ is the sound speed.

When a NS spin-downs enough for the condition $\R{c} > \R{A}$ to be met,
a star leaves the Propeller stage and switches to the next stage, depending
 on the relation between $\R{A}$ and $\R{G}$.
Note, that in our scenario this can also happen because of changes
in the ISM density or in the velocity of a NS.

If $\R{A} > \R{G}$, then a NS enters the Georotator stage, which is called 
so because of similarity of the NS magnetosphere structure to the Earth
magnetosphere in the fast solar wind.
At this stage in our model no spin-up/spin-down mechanisms are taken into account.

Otherwise, if $\R{A} < \R{G}$ then a star at first enters the
subsonic Propeller stage,
where the main accretion mechanism -- the Rayleigh-Taylor instability --
is supressed by  high temperature of an envelope
(i.e., the gas is ``too light'').
Temperature increases because of heating
(a NS loses angular momentum to the envelope and heats it),
and decreases because of radiative losses (mainly bremsstrahlung).
The spin-down rate at this stage is taken in the following form \citep{daviespringle1981}: 
\begin{equation}
\frac{\D P}{\D t} = 2.4 \times 10^{-11} \mu_{30}^2\, m^{-1}\ \mathrm{s\ s^{-1}}.
\label{ssP_spindown}
\end{equation}
Note, that $\dot P$ is constant if the field is not decaying.

 A NS starts to accrete from the ISM and enters the Accretor stage
when heating of an envelope due to rotational energy losses by a NS
becomes less effective than bremsstrahlung cooling,
and so the matter at the magnetospheric boundary becomes so heavy
that the Rayleigh-Taylor instability develops.
This occurs at the critical period \citep{ikhsanov2001}:
\begin{equation}
\Period{break}=8.7\times10^4\ R_\mr{A,\,10}^{5/2}\,\mu_{30}^{-2/3}\,m^{1/6}\ \mr{s}
\end{equation}
where $R_\mr{A,\,10}$ is the Alfven radius 
in units of $10^{10}$ cm. 
At the Accretor stage the spin-down is calculated as at the subsonic Propeller stage
(eq. \ref{ssP_spindown}), and we neglect possible quasi equilibrium at this stage.
This is an oversimplification \citep{ppk02}, 
but as here we are not interested in details of spin properties of Accretors
we neglect some details.

\subsubsection{Magnetic field distribution and decay. Initial period distribution}

We consider two different field distributions.
The first is a delta-function. As a standard value we use $\mu_{30} = 1$,
which gives us $B_\mr{eq} = 10^{12}$~G, according to  $\mu = B_\mr{eq}R^3$.
This distribution is chosen to make comparison with the results from Paper I \nocite{census2000}.

The second distribution is a result of the magnetic field decay starting with the ``optimal'' one from the paper 
\cite{popov10}. This ``optimal'' distribution is the lognormal one with $\langle \log (B_\mr{pole}/[G])\rangle\ = 13.25$
and $\sigma_{\log B_\mr{pole}} = 0.6$, where $B_\mr{pole}$ is the value of the poloidal field on the magnetic pole
($B_\mr{pole} = 2 B_\mr{eq}$).
Till the magnetic field reaches some saturation value $B_\mathrm{min}$ it undergoes decay according to 
\begin{equation}
B(t) = B_0 \frac{ e^{-t/\tau_\mr{Ohm} } }{ 1 + \frac{ \tau_\mr{Ohm} }{ \tau_\mr{Hall} } \left( 1 - e^{-t/\tau_\mr{Ohm} }\right) }
\end{equation}
where $\tau_\mr{Ohm}$ is the Ohmic characteristic time, and $\tau_\mr{Hall}$ is the
typical timescale of the fast, initial Hall stage, that depends on the initial
field ($\tau_\mr{Hall} \propto 1/B_0$). Typically, $\tau_\mr{Ohm} = 10^6$
yrs and $\tau_\mr{Hall} = 10^4$ yrs (for $B_0 = 10^{15}$ G).
The asymptotic
value of the field depends of the initial strength.
In order to approximate the results of simulations by \cite{pons2009} we choose
\begin{equation}
\label{B_min}
B_\mr{min} = \min\left\{\frac{B_0}{2}, 2\times10^{13} \mr{G} \right\}
\end{equation}

Since, according to eq. (\ref{time-spent-as-ejector})
$\min(t_\mr{E}) \approx 10^7$ yr $\gg \tau_\mr{Ohm} = 10^6$ yr, we
neglect the process of decay and take the
field already decayed down to $B_\mr{min}$ as the inital field in our model.
The shape of this distribution is shown in Fig. \ref{results: magnet}.

Below we will refer to these two distributions as ``the standard'' (for $\mu_{30}=1$) and ``the decayed'' respectively.
Everywhere below $B$ is the polar field.

For $P_0$ in the complete model we take the distribution with
$\langle P_0 \rangle = 0.25$~s and $\sigma_\mr{P_0} = 0.1$~s.
Such distribution was used in \cite{popov10}.
In our scenario here this is a simplification, as a NS can
spin-down while its field decays down to the minimum value. So, an
initial period in our model should be different (longer, depending on
the strength of the initial magnetic field) from a period used in
\cite{popov10}. This results in slight overestimating of the number of objects at the Ejector stage
on the price of other stages.
But we tested that this assumption does not influence our results significantly
even for magnetar-scale initial fields. This is so because a star
always evolves
off the Ejector stage much slower than the field decays.

\subsubsection{Velocity distribution}

A NS initial velocity (and so, its kick) has great impact on its magneto-rotational
evolution. This is due to the fact that the efficient accretion rate $\dot M$ 
strongly depends on the velocity:  $\dot M\sim v_\mr{rel}^{-3}$. Note, that $\dot M$ (i.e. just a combination of the ISM density,
relative velocity and NS mass) can be defined for any evolutionary
stage, and it just demonstrates how efficiently a NS interacts with the surrounding medium. 
Almost all characteristic radii and critical periods depend on $\dot M$.

Not only the absolute value of a kick is important, its direction is significant, too.
If a kick has large component perpendicular to the Galactic disc, $v_\mr{z}$,
then a  star spends much less time close to the Galactic plane, where the ISM density is higher (see below).

Most of Accretors in our scenario have small velocities, so only
low-velocity end of the distribution is important.
Several shapes of the kick velocity distribution have been discussed in the literature:
\cite{acc2002, hobbs2005, fgk2006}.
Here we use the initial velocity distribution proposed by \cite{acc2002}.
It is a bi-maxwellian distribution with 
the Gaussian three-dimensional dispersions $\sigma_1= 90\ \kmps$ and $\sigma_2= 500\ \kmps$.
We vary the contribution of each of these components using the parameter $w_1$,
which is the fraction of NSs in the low-velocity component.

\subsection{Semianalytical model}

In a simple semianalytical model we assume monotonic magneto-rotational evolution of a NS in constant conditions:
NS velocity, its magnetic field and the ISM density do not change during a NS lifetime.
Each transition from stage to stage is defined  by solving appropriate equations for the given characterisic values.
In  this  model we always use the NS mass $M = 1.4 M_{\sun}$ ($m = 1.4$) and the
initial period $P_0 = 0.02$~s. 

At first, we solve equations 
$R_\mr{sh}(P, \mu, v) = \R{G}$, $R_\mr{sh}(P, \mu, v) = \R{l}(P)$ to find 
the period at which the Ejector stage ends and the star becomes a Propeller.
This gives us: 
\begin{equation}
P(\mr{E \stackrel{G}{\longrightarrow} P}) \approx 7 \, \mu_{30}^{1/2} v_{10}^{1/2} n^{-1/4} m^{-1/2} \ \mr{s} \\
\end{equation}
for $R_\mr{sh} = \R{G}$ and
\begin{equation}
P(\mr{E \stackrel{LC}{\longrightarrow} P}) \approx 142 \, \mu_{30}^{1/3} v_{10}^{-1/3} n^{-1/6}\ \mr{s}
\end{equation}
for $R_\mr{sh} = \R{l}$.
The critical period (for any branch of the transition from the Ejector stage) is called
below \Period{E}

Integrating eq. (\ref{magneto-dipole-losses}) we obtain:
\begin{equation}
P(t) = \sqrt{ P_0^2 + \frac{16\pi^2\mu^2}{3c^3I} t }\ \ \mr{s}.
\end{equation}
Then, neglecting $P_0$ 
we derive the time of the first transition: 
\begin{equation}
t(\mr{E \stackrel{G}{\longrightarrow} P}) \approx
8.25 \times 10^8 \mu_{30}^{-1} v_{10} n^{-1/2} m^{-1}\ \mr{yrs},
\label{time-spent-as-ejector}
\end{equation}
\begin{equation}
t(\mr{E \stackrel{LC}{\longrightarrow} P}) \approx
3.26 \times 10^{11} \mu_{30}^{-4/3} v_{10}^{-2/3} n^{-1/3}\ \mr{yrs}.
\end{equation}
Among these values the smaller one is used.

If for a given $\mu$ and $v$ the value of $\T{E}$ exceeds $\T{Gal} = 10^{10}$ yr,
then the fraction of lifetime which a star spends as an Ejector, $\w{E}$, is equal to 1.
Elsewhere, $\w{E}$ is equal to $\T{E} / \T{Gal}$.

Next, we must consider the Propeller stage.
Solving the equation $\R{c}(P) = \R{A}(\mu, v)$ for both -- $\R{A} > \R{G}$ and $\R{A} < \R{G}$ -- cases, we obtain 
the critical periods:
\begin{equation}
\Period{}(\mr{P \stackrel{A < G}{\longrightarrow}\ ssP }) =
500\ \mu_{30}^{6/7} v_{10}^{9/7} n^{-3/7} m^{-11/7}\ \mr{s},
\end{equation}
\begin{equation}
\Period{}(\mr{P \stackrel{A > G}{\longrightarrow}\ G }) =
3\times10^5\ \mu_{30}^{1/2} v_{10}^{-1/2} n^{-1/4} m^{-1/2}\ \mr{s}.
\end{equation}
These periods correspond to the end of the Propeller stage.
Any of such periods is called below \Period{P} (note, that before in
several papers such a period was called \Period{A}, as without the
subsonic Propeller stage and neglecting the possibility that a star becomes
a Georotator, after reaching \Period{P} accretion starts).

 Then by solving the equation $\log(\Period{A}) - \log(\Period{E}) = K \Delta t$
on $\Delta t$
we obtain \footnote{
\Period{E} is taken as $P(\mr{E \stackrel{G}{\longrightarrow} P})$ for the case $\R{A} < \R{G}$, 
and $P(\mr{E \stackrel{LC}{\longrightarrow} P})$ -- for the opposite case.
}
the value for  $\Delta \T{P}$. This is the time period during which a NS stays at the Propeller stage.
The fraction of Propellers in the total distribution among evolutionary stages
is taken in the form $\w{P} = \min(1 - \w{E},\ \Delta \T{P}/\T{Gal})$.

If $\R{G}(v) < \R{A}(\mu, v) \leq \R{c}(\Period{P})$ 
a star is considered as a  Georotator.
The fraction of NSs at this stage is $\w{G} = 1 - \left(\w{E} + \w{P}\right)$.

If $\R{A}(\mu, v) < \R{G}(v)$, and due to
the spin-down at the Propeller stage $\R{A}(\mu, v) \leq \R{c}(\Period{A})$, then
the subsonic Propeller stage begins. The spin-down during this stage is given by eq. (\ref{ssP_spindown}).
The stage ends when $P = \Period{break}$,
which gives us the equation: $\Delta t = \left(\Period{break} - \Period{A}\right) / \dot{P}$, where $\dot{P}$ is taken according to eq. (\ref{ssP_spindown}),
from which we obtain $\Delta \T{ssP}$ -- the duration of the
 subsonic Propeller stage.
Fraction of this stage is $\w{ssP} = \min(1 - \left(\w{E} + \w{P}\right), \Delta
\T{ssP}/\T{Gal})$

After a NS spin-downs and leaves the subsonic Propeller stage, it starts to accrete.
The fraction of Accretors is $\w{A} = 1 - \left(\w{E} + \w{P} + \w{ssP}\right)$.

For the scenario without the subsonic Propeller stage we assumed $\w{ssP} = 0$ 
(or, equivalenlty, taking $\w{A} = 1 - \left(\w{E} + \w{P}\right)$, as it was done in Paper I).

\subsection{Complete numerical model}

 Using the same subroutines for characteristic values and spin-down rates
as in the semianalytical model, we make a more detailed numerical model.
A NS evolution in this model  proceeds in  realistic conditions: the Galactic
potential and the ISM density distribution.

 This is done by splitting all the time from 0 to $t_\mr{Gal}$ into an equdistant grid.
Then, on the domain of the acquired time grid we compute all values which do not independ on
the rotation of a star: velocities, coordinates, ISM densities, \R{A},
$\dot M$,  and so on.

 After all rotation-independent values are fetched, the magneto-rotational evolution is calculated
starting with the Ejector stage. Transitions between stages are determined by changes
in relations between characteristic values, as described above.
Note, that there is almost no prohibited transitions, because of large variations
in the environment and in the velocity of a NS.

\subsubsection{Spatial evolution}

 Magneto-rotational evolution strongly depends on the spatial evolution of a NS,
moving through the Galaxy. A NS recieves an initial kick velocity.
The vectors of the kick and the progenitor's Keplerian velocities are summed.
In our model, the ISM rotates with a Keplerian velocity around  the center of  the Galaxy.

Equations of motion of a NS with a given initial values are  solved on the
time grid using the LSODA subroutine \citep{lsoda}.
The Galactic potential is taken   in the same form as in Paper I
(i.e. in the form suggested in  \citealt{mn75} and Pac\'zynski 1990\nocite{p1990}). It is a three component potential 
(disc, buldge, and halo) which reproduces well enough  trajectories on a long time scale (billions of years).
In the problem concidered here we do not need a more complicated gravitational potential. 

\subsubsection{ISM density and NS initial spatial distribution}

The ISM density is taken according to the old analytical model in \cite{posselt08}. 
It is generally the same as in Paper I with some corrections in the z-dependence (see below).
This distribution have exponential or Guassian behaviour perpendicular
to the Galactic disc. The radial distribution has a peak
at $R \sim 5\ \mr{kpc}$. 
For very low-density regions (large $R$ and $z$) we used the minimum value of the ISM number denstity 
$n=10^{-5}$~cm$^{-3}$.

In Paper I there was a small mistake (copied from  \citealt{zane1995}) regarding the ISM distribution.
Dispersions in eqs. (5) and (7) of Paper I
should be divided by 2.35 (as they are actually not dispersions, but FWHM). 
In eq. (6) of Paper I the coefficients  0.345, 0.107, and 0.064
should be  0.7, 0.19, and 0.11, correspondently.

Exactly as in Paper I, the birthrate of NSs is proportional  to the square of the local ISM density. 

\section{Results}

\subsection{Semianalytical model results}

 With this model we address two main questions. How does the account for the subsonic Propeller stage influence
the fraction of Accretors? How does the fraction of accretors depend on the magnetic field?  

In Paper I the subsonic Propeller stage was not used, and only Crab-like
fields, $\sim 10^{12}$~G, were considered. 
Here we include this stage and take into account higher magnetic fields (up to the values typical
for decayed fields of magnetars).
Obviously, the first effect reduces the number of Accretors, while the second -- increases. 
It is interesting to understand with a simple model the interplay between them before addressing
the same questions with a more advanced one. In this model we use only
the delta-function magnetic field distribution.

\begin{figure}
\centering
\includegraphics[width=\columnwidth]{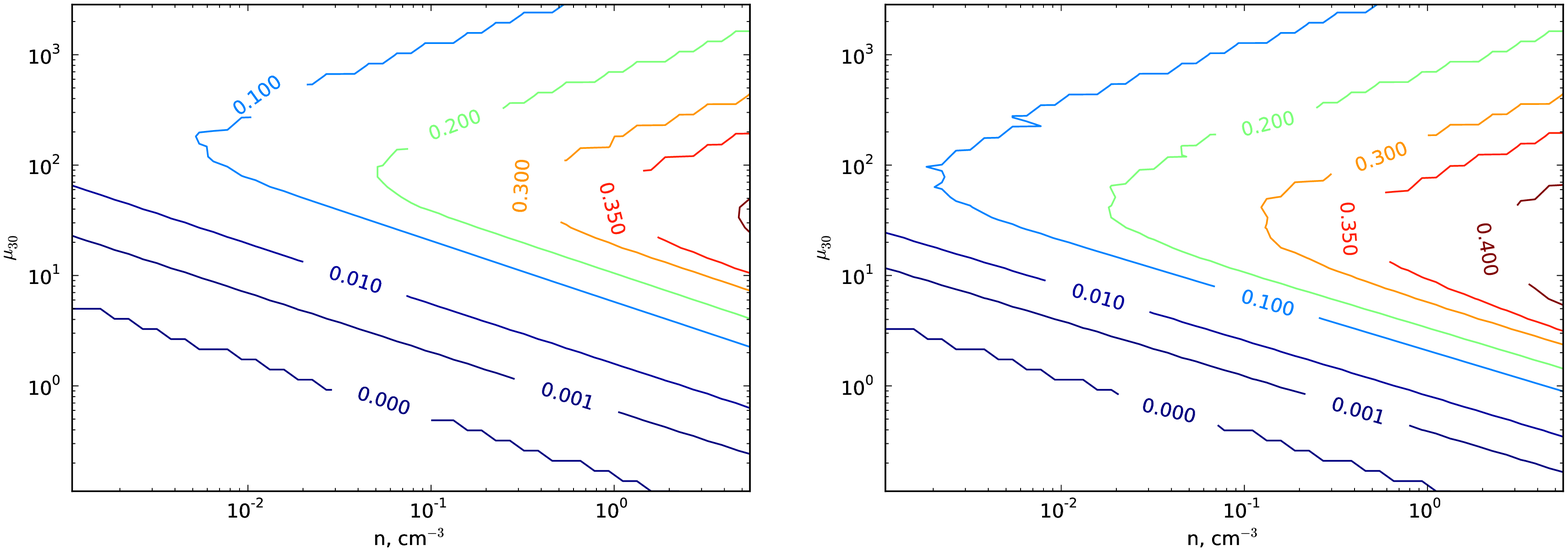}
\caption{Fraction of Accretors in the semianalytical model. The bi-maxwellian kick velocity
distribution from Arzoumanian, Chernoff and Cordes (2002) is taken. 
In the left panel we show results for the scenario with the subsonic Propeller stage.
In the right -- we neglect this stage to demonstrate its influence on our results.}
\label{results:analytical}
\end{figure}

 In Fig. 1 we show the dependence of the fraction of INSs at the Accretor stage on the magnetic field and the ISM number density.
Here both of these parameters are constant during a NS lifetime.
For Crab-like fields and the ISM density $\sim 0.1$~--~1~cm$^{-3}$
it is below $\sim$1\%.
The number of Accretors for a given realistic ISM density steadily grows with increasing initial magnetic fields up to 
$\sim 5 \times 10^{13}$~G. This is related to the fact that a NS with larger magnetic field spin-downs faster, and so quicker 
reaches the stage of accretion. However, for larger fields it happens at large periods, but the increase of the spin-down
rate is more important than the increase of critical periods for transitions.

For large fields the fraction of Accretors starts to decrease, because more and more sources appear as Georotators
due to their huge magnetospheres. However, it is important to repeat, that here we assume constant fields, but it is
normally accepted that large fields gradualy decay while a NS is aging.

The difference with Paper I, where for $\mu\approx 10^{30}$~G cm$^3$, 
$n\approx 1$~cm$^{-3}$ and comparable velocities we obtained $\sim$ several percents of Accretors,
is explained by the influence of the subsonic Propeller stage. It is visible in the right panel.
The fraction of accretors for $n=1$~cm$^{-3}$ and $\mu = 10^{30}$~G cm$^3$
is about one percent if the subsonic Propeller stage is not taken into account.
For the initial field in the M7 range the fraction of Accretors
in our new model (with the subsonic Propeller stage included,
see the left panel) goes up to $\sim 10$~--~30\%.

 Clearly, the main conclusion here is that the subsonic Propeller
is not a strong barrier for NSs with realistically large fields
in contrast with conclusion by \cite{ikhsanov2001}.
As now we know that the fraction of such objects (SGRs, AXPs, M7, RRATs) is not low -- tens of percent, -- one can expect that
significant fraction of old INS can start to accrete.
Thus, one has to study the distribution of INSs in different evolutionary stages in more details.

\begin{table*}
\caption{NS evolutionary tracks}
\begin{center}
\begin{tabular}{l|ccc|cccccc}
\hline
Track & $n$, cm$^{-3}$  & $\mu_{30}$ & $v_{10}$ &
		\w{E} & \Period{E}, s & \w{P} & \Period{P}, s &
		\w{ssP} & \Period{break}, s \\
\hline
Track I & $0.5$ & $1$ & $5$ & $0.419$ & $16.051$ & $0.423$ & $3.163 \times 10^{3}$ & $0.850$ & $2.278 \times 10^{6}$ \\
Track II & $0.5$ & $1$ & $20$ & -- & -- & -- & -- & -- & -- \\
Track III & $0.5$ & $1$ & $40$ & -- & -- & -- & -- & -- & --  \\
Track IV & $0.5$ & $10$ & $5$ & $0.042$ & $50.758$ & $0.042$ & $2.276 \times 10^{4}$ & $0.067$ & $1.317 \times 10^{7}$ \\
Track V & $0.5$ & $10$ & $20$ & $0.168$ & $101.517$ & $0.170$ & $1.353 \times 10^{5}$ & $0.651$ & $2.568 \times 10^{8}$ \\
Track VI & $0.5$ & $10$ & $40$ & $0.163$ & $100.091$ & $0.169$ & $1.523 \times 10^{5}$ & \multicolumn{2}{l}{Georotator} \\
Track VII & $2.0$ & $1$ & $5$ & $0.209$ & $11.350$ & $0.212$ & $1.746 \times 10^{3}$ & $0.370$ & $8.464 \times 10^{5}$ \\
Track VIII & $2.0$ & $1$ & $20$ & $0.838$ & $22.700$ & $0.854$ & $1.038 \times 10^{4}$ & -- & -- \\
Track IX & $2.0$ & $1$ & $40$ & -- & -- & -- & -- & -- & -- \\
Track X & $2.0$ & $10$ & $5$ & $0.021$ & $35.892$ & $0.021$ & $1.257 \times 10^{4}$ & $0.030$ & $4.892 \times 10^{6}$ \\
Track XI & $2.0$ & $10$ & $20$ & $0.084$ & $71.783$ & $0.085$ & $7.469 \times 10^{4}$ & $0.264$ & $9.541 \times 10^{7}$ \\
Track XII & $2.0$ & $10$ & $40$ & $0.103$ & $79.442$ & $0.106$ & $1.077 \times 10^{5}$ & \multicolumn{2}{l}{Georotator} \\
\hline
\end{tabular}
\label{typicaltracks}
\end{center}
\end{table*}

Some results of this subsection are summarized in the Table. 1.
We demonstrate twelve tracks for different $\mu, n, v$. NSs following tracks II, III, and IX always stay at the Ejector stage.
High velocity NSs with fields larger than the Crab-like value, become
Georotators after the Propeller stage (tracks VI and XII).
The NS following track VIII never becomes an Accretor as it stays for a long time at the subsonic Propeller stage. 
All the rest NSs (tracks I, IV, V, VII, X, XI) finally start to accrete.

\subsection{Complete numerical model}

 In this subsection we present our main results obtained with the complete model for  $10^5$ evolutionary tracks.
Our main aim is to calculate the distribution in evolutionary stages for two different  distributions
of the initial magnetic field. In addition, we demonstrate the effect of changing velocity distribution.

\begin{figure}
\centering
\includegraphics[width=\columnwidth]{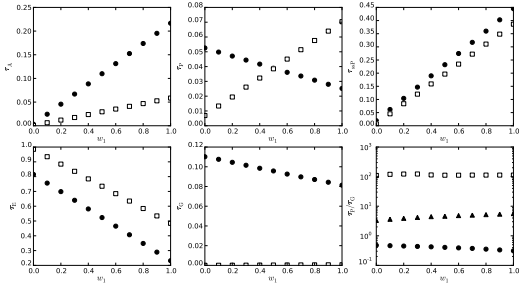}
\caption{Different panels show fractions of INSs at different stages in the complete numerical model:
Accretors (top left), Propellers (top center), subsonic Propellers (top right), Ejectors (bottom left) 
and Georotators (bottom center). Black circles denote the
{\it decayed} magnetic field distribution, whereas empty squares corresponds to the delta-function $\mu=10^{30}$~G~cm$^3$. 
Bottom right panel shows the ratio of Propellers to Georotators. Triangles correspond to an additional variant of
magnetic field distribution.
In this variant only low-field -- $B \le 5\times10^{12}$~G -- INS are taken from the
{\it decayed} model. By $w_1$ we mark the fraction of low-velocity component of the bi-maxwellian
distribution from Arzoumanian, Chernoff and Cordes (2002). 
Here $\mu = BR^3/2$, as $B$ is the magnetic field value at the pole.
}
\label{results:numerical}
\end{figure}

 As we use the bi-Maxwellian distribution proposed by \cite{acc2002}, to demonstrate the dependence of
our results on the velocity distribution, we decided to change relative
contributions of the two constituents. 
In Fig.\ref{results:numerical} the horizontal axis shows $w_1$ -- the contribution of the low-velocity part of
the bi-Maxwellian distribution. For $w_1=0$ we have a pure Maxwellian distribution with $\sigma= 500\ \kmps$, for
$w_1=1$ -- a pure Maxwellian with $\sigma= 90\ \kmps$. 

 We show results of calculations for two distributions of initial magnetic fields described above (sec.2.1.2). 
The first is just a delta-function
$\mu=10^{30}$~G~cm$^3$. It corresponds to the typical assumption made in 90s. The second is based
on recent results by \cite{popov10}. 

 Fractions of Ejectors, Propellers, subsonic Propellers, Accretors and Georotators demonstrate monotonic, nearly linear
behavior. The number of Ejectors strongly decreases with increasing $w_1$. The behavior of Accretors and subsonic
Propellers is opposite. 

 The behavior for the two studied field distributions is similar in the cases of Accretors, Ejectors, and subsonic Propellers.
In the case of Propellers and Georotators the situation is different for two distributions. 

 In the most realistic case according to \cite{acc2002}, -- $w_1=0.4$ -- 
we have (in the case of initially decayed field distribution) $\sim 55$\%
of Ejectors, $\sim 5$\% of supersonic and $\sim 20$\% of subsonic Propellers, $\sim 10$\% of Accretors, and finally, 
$\sim 10$\% of Georotators. As we see, now for our ``the best choice'' model we predict more Accretors than in Paper I.
It is what was expected on the basis of the semianalytical model. 

\begin{figure}
\centering
\includegraphics[width=0.5\textwidth]{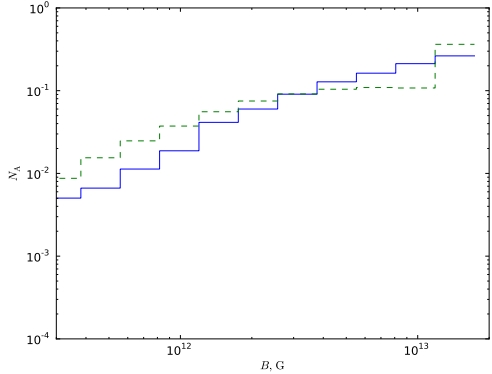}
\caption{The dashed line represents the field distribution after
the immediate field decay discussed in sec. 2.1.2, 
i.e. the {\it decayed} model. The solid line shows
the contribution to the number of Accretors for different initial fields in the decayed model, normalized to unity.
The velocity distribution is calculated for $w_1=0.4$.
Here $B$ is the magnetic field value at the magnetic pole. 
Each bin shows mean fraction of time, among all the stars having such a fields.}
\label{results: magnet}
\end{figure}

The increase in the relative number of Accretors is due to the presence of INSs with large initial magnetic fields. 
This is illustrated in 
Fig.\ref{results: magnet}. We show there contributions of INSs with different initial magnetic fields to the population
of Accretors. Note, that the scale is logarithmic in both axis. 
INSs with initial fields $< 3\,10^{12}$~G are more numerous than those with $10^{13}$~G $< B < 2\,10^{13}$~G.\footnote{Here 
we speak about decayed fields in our model. If we recalculate it to obtain real pre-decay initial fields, then the intervals
are changed.}
However, the latter produce seven times more Accretors.
Still, many (about 1/2)
 of NSs with the largest initial field considered here do not produce many Accretors as they become Georotators due to
large spatial velocities.

Typically, NSs become Accretors in the regions of the Galactic disc, where the ISM density is higher. 
As it was noted before NSs with low total velocity but significant z-component,
$v_\mr{z}$, spend most of its lifetime outside the Galactic plane.
Most likely the longest stage for such a NS is the subsonic Propeller.
If a star is born relatively far from the Galactic center and recieves  a large kick 
then it escapes from the Galaxy. Such a NS spends most of its life
as an Ejector. Alternatively, if the velocity is high but not enough 
to escape the Galaxy, the NS returns to the Galactic plane after some long time,
then it can quickly  pass the Propeller stage
(so-called {\it non-gravitating Propeller} in this case)
and become a Georotator.

 In addition to the global distribution we compute separately distributions over stages inside ($R < 16$ kpc and
$|z| < 1$ kpc) and outside the Galaxy (in the following paragraphs we refer as ``the Galaxy'' only to the former volume).
It can seem suprising and confusing, but according to our model we predict more
Accretors than Ejectors inside the Galaxy: $\w{A} \sim 30 \%$ and
$\w{E} \approx 20 \%$.
Subsonic Propellers are more abundant than Accretors and Ejectors
in this volume:  $\w{ssP} \sim 43 \%$.
These numbers can be explained in the following way. 
Most of NSs which contribute a lot to the number of Accretors
have $v_\mathrm{z} < 100\ \kmps$.
They  spend most of their lives inside the Galaxy, and so there they dominate.
Other stages, correspondently, are not abundunt inside the Galaxy:
Georotators contribute $\w{G} \sim 7\%$, Propellers -- $\w{P} \sim 1-2\%$.
Roughly, NSs with kick velocities from the low-velocity part of the distribution (about 30-40 \%)
stay inside the Galaxy. 
Those with magnetic fields higher than typical radio pulsar values become Accretors. 

 The situation outside the  cylinder $R < 16$ kpc and
$|z| < 1$ kpc is the following.  
Ejectors contribute $\w{E} = 76\%$, Georotators -- $\w{G} \approx 11\%$, subsonic Propellers -- $\w{ssP} = 9\%$,
Propellers -- $\w{P} \approx 4\%$, and Accretors -- $\w{A} \approx 0.2\%$.
As one can see, the situation with Ejectors and Accretors is opposite in comparison with the internal part.
Almost all Accretors are situated  inside the Galaxy.
Note, that due to this Fig. \ref{results: magnet} with the total distribution
also refers to the population of Accretors inside the Galaxy.
Almost $2/3$ of all neutron stars are outside the Galaxy
and they are either Ejectors or Georotators.

 In the solar neighborhood ($R_\mathrm{solar} < 2$ kpc and $|z| < 0.5$ kpc)
we predict $\sim 35-40\%$ of Accretors and slightly more ($\sim 40-45\%$)
subsonic Propellers with only $\sim 18-20\%$ of Ejectors. Contributions of others
stages are negligible. In total, in the solar
proximity ($R_\mathrm{solar} < 2$ kpc and $|z| < 0.5$ kpc)
there are $0.33\%$ of all NSs. This gives us,
for $N_\mathrm{NS} = 10^9$ the number density in the solar neighborhood
$n_0 \approx 3 \times 10^{-4}$ pc$^{-3}$, in good correspondence with recent results by
\cite{ofek2009} and with earlier studies.

\section{Discussion}

Available estimates of the number of Accretors should be taken with care, since there are several effects
which act in inhibiting accretion, as it is discussed below.
I.e., the number of {\it observable} Accretors can be much smaller.
Still, low velocity INSs moving through a high density medium (the only ones with high enough luminosities
to be potentially detected) and with strong magnetic fields can become accretors after $\la $~few Gyr. It was shown above
that INSs like the M7, which have magnetic fields higher
that those typical of radio pulsars, and at the same time have not very large spatial velocities,
are the most favored as Accretors predecessors.

Although the number of AINSs might be even larger than that originally estimated in
Paper I, the conclusion that at low fluxes Accretors outnumber cooling isolated NS (Coolers) is based on
the assumption that the luminosity corresponds to the Bondi accretion rate. This is a quite
controversial issue. \cite{blaes95} have shown that for typical ISM densities ($n\approx
1\ {\rm cm^{-3}}$) accretion rate does not exceed $\sim$ few $10^9\ {\rm g\, s^{-1}}$, even if the star velocity
drops below $\sim 60\ {\rm km\, s^{-1}}$. This is due to the ionization of the ISM surrounding the
star by the X-ray radiation which, in turn, produces an increase in the sound speed freezing the accretion
rate.
However, in Paper I  it was shown that the velocity distribution of Accretors peaks
at $\sim 50\ {\rm km\, s^{-1}}$, and for these velocities the effect is small. So, for most
Accretors, heating of the ISM can be neglected, especially if they appear in regions of high ISM density.
A further issue is the role played by the star magnetic field in the accretion flow dynamics outside
the Propeller stage. On the basis of 2D MHD calculations, \cite{toropina03} concluded that only a
fraction of the initial (Bondi) flow reaches the star surface, and this fraction decreases with growing
magnetic field of a NS. Whether 3D instabilities may counteract this effect is still an open question.

Anyway, if several weak sources without measured proper motions and
interpreted as INS candidates can be identified in ROSAT, Chandra or/and XMM-Newton archives (see \citealt{turner2010} and
references therein), or 
discovered by eROSITA, then it is not trivial to distinguish Coolers from Accretors.

If an INS comes to the Accretor stage only after a long subsonic Propeller episode, its spin period is
$\ga 10^4$~s.
Such long periods are not unexpected even if the subsonic Propeller stage
is neglected. When a NS starts to accrete it continues to spin down, until it reaches a
quasi-equilibrium period, $P_\mr{{eq}}\approx 10^6$~s for $n=1$~cm$^{-3}$
\citep{kp97,ppk02}. The ultra-long spin periods of Accretors could be the best
discriminator between this type of sources and Coolers, which are expected to have spin periods $\la$~few seconds
(like the M7 and cooling PSRs). 
However, at low fluxes it would
be extremely difficult to discover pulsations in Coolers, so the non-detection
of a periodicity is not a strong argument in favour of an AINS.

Opposite to Coolers, Accretors are expected to show both -- spin-up and spin-down -- as their periods fluctuate around the
quasi-equilibrium value. However, $\dot P$ measurements can be impossible for faint sources with very long periods.

The period of accreting INSs can be significantly shorter
than $\approx 10^6$~s in the case of magnetic fields decaying down to small values ($\sim 10^9$~G), although some kind of
fine tuning is necessary. As discussed above, to reach accretion in a time shorter than the
Hubble time an INS should have at least a magnetic field $\approx 10^{12}$ G. So, decay should not be
significant during the first $\sim$~1 Gyr of the evolution, otherwise a NS spends all its life as an Ejector or
a Propeller \citep{c98,l98,pp00}. If the field decays during the Accretor (or even subsonic Propeller)
phase, an INS can attain a period $\sim10^3$--$10^4$~s, since  $P_\mr{{eq}}$ is
smaller for smaller fields.

Accretors, at variance with coolers, are not expected to be steady sources
because of changes in the accretion rate, due to inhomogeneities of the
ISM, on a time-scale
\begin{equation}
\label{time}
t \approx \frac{R_\mr{G}}{v}\sim 3\times 10^8 v_{10}^{-3}\ {\rm s}\, .
\end{equation}
Note that this time scale is shorter for fainter
sources. 

Spatial distribution of Accretors and new weaker Coolers are expected to be slightly different,
as the first represent much older population, and for the first higher ISM density is favorable for detection in contrast
with the second. New (i.e., undiscovered, yet)
Coolers according to \cite{posselt08} are expected to be found at distances $\sim 1 $~kpc. So, they 
should be relatively bright, $\sim 10^{31}$~erg~s$^{-1}$. Accretors cannot be that bright, and so they are expected to be
found closer. Young Coolers should trace starforming regions. Accretors, which already experienced long evolution in the Galactic 
potential, should be distributed more smoothly. However, for them to be detectable it is important 
to be inside regions of relatively high ISM density.

The X-ray spectrum of a NS accreting at low rate from the ISM is very
similar to those of cooling INSs, at least in the case when the latter has a H atmosphere
(\citealt{tr00} and references therein). Nevertheless, for the same luminosity, the
effective temperature of an Accretor is higher and, hence, the spectrum is harder because of
significantly reduced emitting area. For typical values of the star mass and radius, the
hot polar cap size is

\begin{equation}
\label{rcap}
R_\mr{cap}\sim 9.5\times 10^3 \mu_{30}^{-2/7} v_{10}^{-3/7} n^{1/7}\ {\rm cm}\, .
\end{equation}
This is smaller than the size of a typical emitting area in Coolers. Spectra of Accretors are expected to be harder than those
of Coolers.

We summarize some differences between Accretors and cooling NSs (Coolers) in Table 2.

\begin{table*}
\caption{Comparison of properties of dim Accretors and Coolers}
\begin{center}
\begin{tabular}{lcc}
\hline
 & Accretors & Coolers\\
Spectrum & Harder, $\sim $~hundreds eV & Softer, $\sim 100$~eV \\
Spin periods & Very long, $>10^5$~s & Shorter, $\sim10$~s\\
$\dot p$ & Variable  & Stable spin-down\\
Distance & Close, $\sim 100$~--~200~pc& Further away, $\sim 1$~kpc\\
Luminosity & Low, $\sim 10^{29}$~erg~s$^{-1}$& Higher, $\sim 10^{31}$~erg~s$^{-1}$  \\
Variability & Variable, $\Delta t \sim $~weeks~--~yrs & Stable\\
Spatial distribution & Towards higher gas density & Towards starforming regions\\
\hline
\label{accvscoo}
\end{tabular}
\end{center}
\end{table*}


\section{Conclusions}

After the first of the M7  have been discovered \citep{walter96}, several authors proposed and discussed
that they can be AINSs \citep{walter96, kp97, nt99}.
Though, it appeared that it is not so. The M7 are
young NSs with relatively large fields. Probably, they are related to evolved magnetars \citep{popov10}. 
Here we demonstrate that in future the M7 and similar sources are expected to become AINS if 
their magnetic fields do not decay significantly. Even a relatively long stage of subsonic Propeller \citep{ikhsanov2001} cannot
prevent accretion. This is a good news for observers. Probably, telescopes like eROSITA aboard Spektr-RG will be able
to detect AINS, soon. However, the question of the accretion efficiency is still on the list \citep{toropina03}. 

The distribution over evolutionary stages strongly depends on kick velocity distribution, initial magnetic field distribution
and field evolution. Because of that precise predictions are not possible now. This shows how important is to detect old
isolated NSs as Accretors (or, less probable, other stages) to learn more about initial properties and evotuion of INSs.

\section*{Acknowledgments} 
S.P. thanks Profs. Jose Pons, Aldo Treves, and Roberto Turolla for discussions. 
This work was supported by the RFBR grants 07-02-00961 and 09-02-00032, and by the Federal program for scientific and educational personnel.

\bibliography{boldin}

\end{document}